\documentclass[sigconf,nonacm]{acmart}

\usepackage{booktabs}
\usepackage{graphicx}
\usepackage{xcolor}
\usepackage{multirow}
\usepackage{multicol}
\usepackage{letltxmacro}

\begin{document}

\title{What Gets Hacked in a Collaborative Reward for LLM Recommenders: The Part That Does Not Depend on the User}


\author{Zhuangsheng Li}
\authornote{The first three authors contributed equally to this work.}
\email{zhl214@ucsd.edu}
\affiliation{%
  \institution{University of California, San Diego}
  \city{La Jolla}
  \state{CA}
  \country{USA}
}

\author{Liang Dong}
\authornotemark[1]
\email{lidong@google.com}
\affiliation{%
  \institution{Google Inc.}
  \city{Mountain View}
  \state{CA}
  \country{USA}
}

\author{Yu Wu}
\authornotemark[1]
\email{yuw332@ucsd.edu}
\affiliation{%
  \institution{University of California, San Diego}
  \city{La Jolla}
  \state{CA}
  \country{USA}
}

\author{Dailin Li}
\email{esd4uq@virginia.edu}
\affiliation{%
  \institution{University of Virginia}
  \city{Charlottesville}
  \state{VA}
  \country{USA}
}

\author{Yu Xia}
\email{yux078@ucsd.edu}
\affiliation{%
  \institution{University of California, San Diego}
  \city{La Jolla}
  \state{CA}
  \country{USA}
}

\author{Yunhan C. Zhang}
\email{yunhan3@student.unimelb.edu.au}
\affiliation{%
  \institution{University of Melbourne}
  \city{Melbourne}
  \state{VIC}
  \country{Australia}
}

\author{Rui Cai}
\email{ruicai@ucdavis.edu}
\affiliation{%
  \institution{University of California, Davis}
  \city{Davis}
  \state{CA}
  \country{USA}
}

\author{Yi Wu}
\email{wuyish@google.com}
\affiliation{%
  \institution{Google Inc.}
  \city{Mountain View}
  \state{CA}
  \country{USA}
}

\author{Julian McAuley}
\email{jmcauley@ucsd.edu}
\affiliation{%
  \institution{University of California, San Diego}
  \city{La Jolla}
  \state{CA}
  \country{USA}
}

\renewcommand{\shortauthors}{Li et al.}

\begin{abstract}
Collaborative-filtering (CF) scores provide graded, personalized rewards for training large language model (LLM) recommenders through reinforcement learning. 
However, these rewards can cause exposure collapse: even though the CF model itself produces diverse top recommendations across users, an LLM optimized with its rewards concentrates on fewer items.
We decompose
each score into an \textbf{\textit{item-mean component}}, the item's average
CF score across user histories, and a \textbf{\textit{residual component}},
the current user's score deviation from the average. Through controlled
interventions, we identify the item-mean component as a contributor to
exposure collapse. 
Accordingly, we propose two reward corrections: per-item centering removes each item’s estimated mean score, while per-item standardization further rescales the residual using its estimated standard deviation. Experiments on two datasets with
two identifier schemes show that both corrections substantially improve catalog coverage and exposure evenness. Compared with training on the uncorrected CF reward, our corrections increase the number of distinct items ranked first across users by up to 9.8×, while maintaining comparable ranking accuracy.
Code is available at
\url{https://anonymous.4open.science/r/LLM4Rec-CF-Bias-27EF/}. 
\end{abstract}


\maketitle

\section{Introduction}

Collaborative-filtering (CF) scores provide personalized rewards for training large language model (LLM) recommenders through reinforcement learning (RL). Compared with binary matching rewards, CF scores provide
more fine-grained feedback by distinguishing candidates that do not
match the observed next item. These scores reuse collaborative
information learned from user--item interaction histories~\cite{resnick1994grouplens,sarwar2001itemcf,koren2009mf} without
requiring additional reward annotations. Prior work, including
MiniOneRec and ReRe, has explored and evaluated such
\textbf{\textit{collaborative
rewards}}~\cite{kong2025minionerecopensourceframeworkscaling,tan2025reinforcedpreferenceoptimizationrecommendation}. In this setup, an LLM
policy first undergoes supervised fine-tuning (SFT) on user--item
interaction histories. A frozen CF model then scores each generated
item given the user's history, and RL uses this score to update the
policy.

\begin{figure*}[t]
\centering
\includegraphics[width=0.85\textwidth]{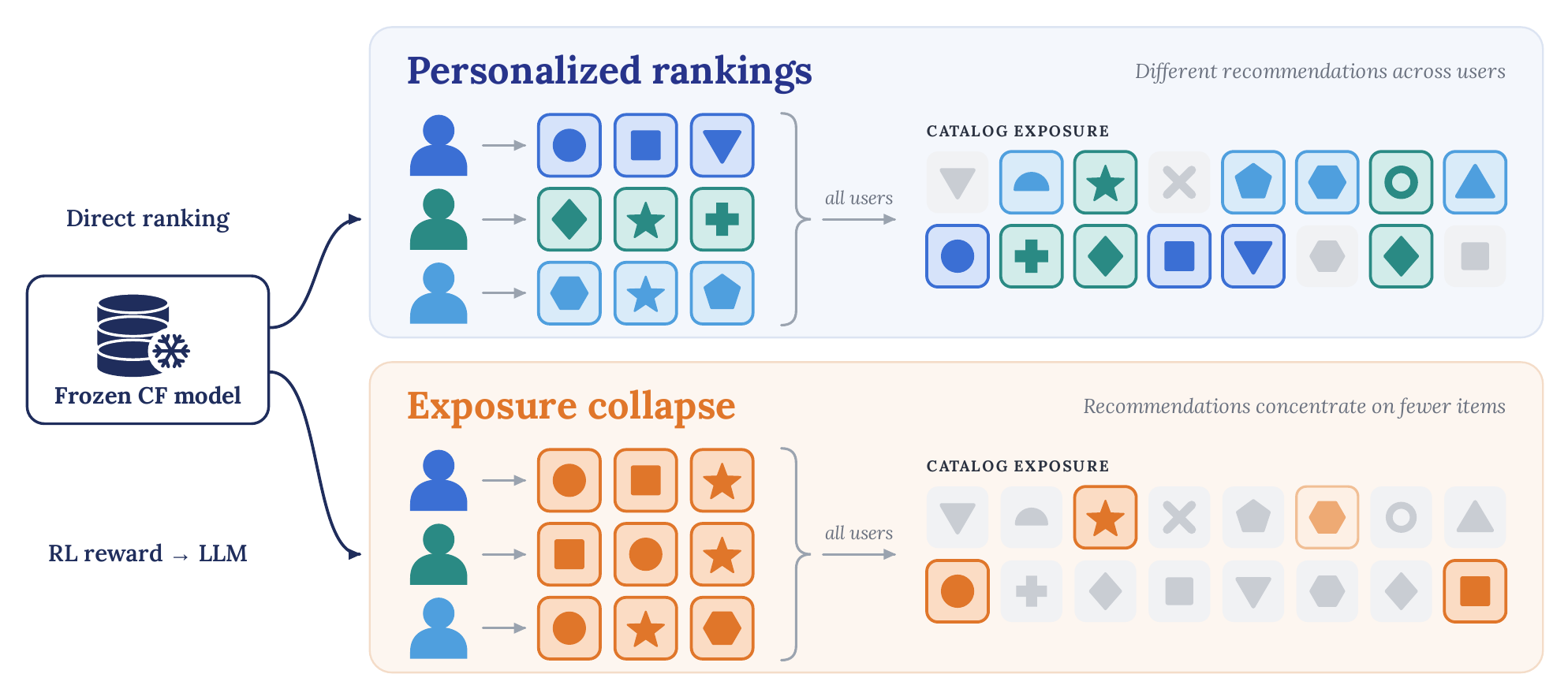}
\caption{Conceptual overview. The frozen CF model produces personalized
rankings when ranking the catalog directly. When its scores are used as
RL rewards, the trained LLM policy's recommendations concentrate on fewer items (\S\ref{sec:prelim} defines
the exposure measures and collapse). Qualitative; no counts depicted.}
\label{fig:teaser}
\end{figure*}

However, using these scores directly as RL rewards can concentrate
recommendations on fewer items across users, even when recommendation
accuracy changes little. For example, on Beauty with Semantic IDs,
HR@10 changes from 0.0851 to 0.0843, while the number of
\textbf{\textit{distinct top-1 items}} ranked first across users falls
from 1,317 to 837 (Table~\ref{tab:controls}). Across Amazon Beauty and
Sports with Semantic and Atomic IDs, this count falls by 36--80\%
relative to the SFT start (Figure~\ref{fig:problem}). These results show
substantial \textbf{\textit{exposure collapse}}:  Direct-ranking controls confirm that the same frozen
CF model can produce user-specific rankings (\S\ref{sub:nothacking}).
This contrast shows that its personalized ranking ability does not
automatically carry over to the LLM policy trained on its scores.

Existing work has also identified problems with CF rewards. MiniOneRec
reports performance degradation when testing a collaborative reward and
hypothesizes reward hacking~\cite{kong2025minionerecopensourceframeworkscaling}; ReRe observes
increasing collaborative rewards alongside declining recommendation
accuracy~\cite{tan2025reinforcedpreferenceoptimizationrecommendation}. However, neither study isolates which
CF score components contribute to these training failures.

To understand the reward's role in exposure collapse, we examine the part
of the CF score shared across users. We decompose each CF score into an
\textbf{\textit{item-mean component}}, the item's mean CF score over a
common, fixed reference distribution of user histories, and a
\textbf{\textit{residual component}}, the current user's score deviation
from that mean (\S\ref{sub:decomp}). Training on estimates of the two
components separately reveals their different effects: on Beauty with
Semantic IDs, the item-mean component alone leaves 32 distinct top-1
items, whereas the residual component alone leaves 1,830
(Table~\ref{tab:controls}). These interventions identify the item-mean
component as a contributor to the policy's exposure collapse.

This diagnosis motivates two per-item reward corrections.
\textbf{\textit{Per-item centering}} subtracts the estimated item-mean
component from each CF score. Since centering leaves each item's score
variance across users unchanged, \textbf{\textit{per-item
standardization}} extends the correction by also normalizing the
residual score scale. Both corrections estimate item statistics once
from a common set of training histories and keep the CF model
frozen during RL (\S\ref{sec:method}).

Across the two datasets and two identifier schemes, both corrections
broaden exposure and improve catalog coverage and exposure evenness
relative to the uncorrected CF reward. Per-item centering and
standardization yield 1.5--4.1 and 1.6--9.8 times as many distinct top-1
items, respectively (\S\ref{sub:fix}). These gains in distinct top-1
items also hold across four CF model architectures and on two additional
LLM backbones on Beauty with Semantic IDs
(\S\ref{sub:scope}).

\begin{samepage}
The main contributions are as follows:
\begin{itemize}
    \item We identify the item-mean component as a contributor to
    exposure collapse through controlled interventions.

    \item We substantially broaden exposure through per-item centering
    and standardization of CF rewards alone.

    \item We validate these gains across datasets,
    identifier schemes, CF architectures, and LLM backbones.
\end{itemize}
\end{samepage}

\section{Related Work}

\subsection{Collaborative Rewards}

Collaborative rewards use scores from pretrained CF models to provide
graded feedback for generated items~\cite{kong2025minionerecopensourceframeworkscaling,tan2025reinforcedpreferenceoptimizationrecommendation}.
MiniOneRec replaces the ranking term with frozen SASRec logits in a
reward ablation, reports degraded accuracy, and discusses possible
reward hacking. ReRe tests collaborative rewards as a dense alternative
and observes increasing rewards alongside declining recommendation
accuracy. Both adopt combinations of rule-based correctness rewards and
ranking-aware penalties in their final designs.

The contribution of individual CF score components to exposure
concentration remains unclear. HiLaR~\cite{hu2026hierarchicallatentreasoningllmbased} standardizes and
clips the CF scores used in its preference reward. Its statistics are
computed across candidate items for a given user, whereas the proposed
corrections estimate
each item's score statistics across a common reference distribution of
user histories. The present study uses controlled component
interventions to identify the item-mean component as a contributor to
cross-user exposure collapse, including cases where the frozen CF
scorer retains personalized rankings when used directly.

\subsection{Popularity Bias}

Popularity bias favors frequently interacted items and can limit
long-tail exposure~\cite{Klimashevskaia_2024,abdollahpouri2019unfairnesspopularitybiasrecommendation,chen2023bias}.
Catalog coverage complements accuracy by measuring the range of
recommended items~\cite{10.1145/1864708.1864761}.

\begingroup
\setlength{\emergencystretch}{1em}
Methods for mitigating this bias intervene at different stages~\cite{zhang2021pda,wei2021macr}.
IFairLRS adjusts instruction-tuning sample weights and reranks outputs
to match item-group exposure to user histories~\cite{jiang2024itemsidefairnesslargelanguage}.
SPRec~\cite{Gao_2025} uses preference training to curb over-recommendation:
it constructs pairs from logged positives
and model-generated negatives, alternating SFT with DPO~\cite{rafailov2023dpo}.
Items assigned higher generation probabilities receive stronger
suppression. At the tokenization and generation stages, Ghost
combines popularity-aware semantic tokenization with asymmetric
unlikelihood training to reduce disadvantages for tail
items~\cite{yin2026echoesfilterbubblediagnosing}. GPTRec combines ranking accuracy and a
weighted penalty on recommended items' mean popularity in the RL
reward~\cite{petrov2024aligninggptrecbeyondaccuracygoals}.
\par\endgroup

Beyond popularity bias, prior work addresses key-term confounding in
conversational recommendation~\cite{10.1145/3580305.3599539}, uses
interventional feedback for LLM debiasing~\cite{xia-etal-2024-aligning},
evaluates user-specific reward preferences~\cite{ma2026personalizedrewardbench},
and diagnoses modality interference through controlled
perturbations~\cite{cai2026modalityinterference}.
The proposed corrections act on item-dependent scoring tendencies in
an already-trained CF model before its scores drive policy updates.
The item-mean component measures the model's average score across user
histories; its association with interaction-count popularity
(Appendix~\ref{app:more}) does not equate the two quantities.

\section{Preliminaries and Problem Formulation}
\label{sec:prelim}

\begin{figure}[t]
\centering
\includegraphics[width=\columnwidth]{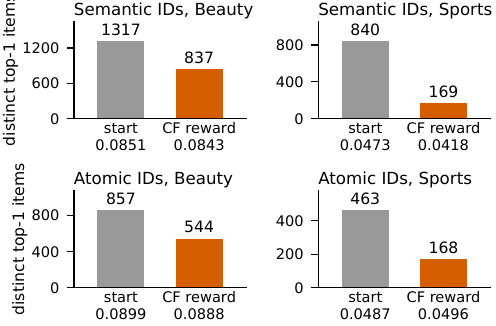}
\caption{Distinct top-1 items at the SFT start and after RL on the
CF reward, under both identifier schemes on Beauty and Sports;
HR@10 below bars; three-seed RL means.}
\label{fig:problem}
\end{figure}

\subsection{LLM-based Sequential Recommendation}

Given a user's interaction history, sequential recommendation predicts
the next item from a fixed catalog $\mathcal I$. An LLM recommender
performs this task by generating an item identifier conditioned on the
history~\cite{geng2023recommendationlanguageprocessingrlp}. Two identifier schemes are considered: \textbf{\textit{Semantic
IDs}} represent items through sequences of content-based
tokens~\cite{rajput2023recommendersystemsgenerativeretrieval,chen2026grounded}, whereas \textbf{\textit{Atomic IDs}} assign a
dedicated token to each item~\cite{Hua_2023}.

The recommendation policy first undergoes supervised fine-tuning (SFT)
on user--item interaction histories and their next-item targets~\cite{ding2026riser}. The
resulting checkpoint, termed the \textbf{\textit{SFT start}}, initializes
subsequent RL training. Within each setting, all RL variants share this
initialization. At evaluation time, the policy generates a ranked list
of valid catalog items, with the top ten forming the recommendation list
$\mathcal R_u$ for user $u$.

\subsection{RL with Collaborative Rewards}

A collaborative-filtering model trained on user--item interactions can
supply personalized feedback for RL post-training. Let
$f_{\mathrm{CF}}(u,i)$ denote its score for item $i$ given user $u$'s
interaction history. The \textbf{\textit{collaborative (CF) reward}} uses
this score directly:
\begin{equation}
r(u,i)=f_{\mathrm{CF}}(u,i).
\label{eq:cf_reward}
\end{equation}
During post-training, the LLM policy generates candidate items, the CF
model scores them, and RL uses these rewards to update the policy. The
CF model remains frozen throughout this process. The setting considered
here uses critic-free policy-gradient training, with
GRPO~\cite{shao2024deepseekmathpushinglimitsmathematical} as the main optimizer.

The distinction between the CF \textbf{\textit{score}} and the training
\textbf{\textit{reward}} identifies the object of intervention. In the
uncorrected setting, termed the \textbf{\textit{CF arm}}, the score is
supplied directly as the reward. Reward corrections instead transform
$r(u,i)$ before it is used by RL, without retraining the CF model
(\S\ref{sec:method}). Thus, the scorer remains fixed while different
rewards can lead the recommendation policy to produce different exposure
patterns.

\subsection{Measuring Exposure Concentration}
\label{sub:metrics}

Exposure concentration is evaluated across a common set of test users
$\mathcal U_{\mathrm{eval}}$. The primary measure counts
\textbf{\textit{distinct top-1 items}}:
\begin{equation}
D_{\mathrm{top1}}
=\left|\left\{\mathrm{Top1}(u):u\in\mathcal U_{\mathrm{eval}}\right\}\right|,
\label{eq:distinct_top1}
\end{equation}
where $\mathrm{Top1}(u)$ is the first item in $\mathcal R_u$. This count
measures how broadly first recommendations are distributed across the
catalog~\cite{10.1145/1864708.1864761}.

Two complementary metrics characterize exposure within the full top-ten
lists. The exposure of item $i$ is
\begin{equation}
e(i)=\sum_{u\in\mathcal U_{\mathrm{eval}}}\mathbf{1}[i\in\mathcal R_u].
\label{eq:exposure}
\end{equation}
\textbf{\textit{Coverage@10}} measures the fraction of catalog items with
nonzero exposure~\cite{10.1145/1864708.1864761}, while the \textbf{\textit{Gini
coefficient}} of $e$ measures exposure inequality over the full catalog,
including items with zero exposure~\cite{doi:10.1287/mnsc.1080.0974}. Higher
coverage indicates broader exposure, and a lower Gini coefficient
indicates a more even distribution. We also evaluate recommendation
accuracy using HR@10 and NDCG@10.

Exposure changes induced by CF-reward training are measured relative to
the SFT start. This study uses a reduction of at least 30\% in distinct
top-1 items as an operational criterion for \textbf{\textit{exposure
collapse}}. The effectiveness of reward correction is evaluated relative
to the uncorrected CF arm, using changes in distinct top-1 items,
coverage, and Gini.

The research problem is to identify which components of the frozen CF
model's scores contribute to exposure concentration during RL
post-training, and whether correcting these rewards can broaden
exposure without retraining the scorer.

\section{Per-Item Reward Correction}
\label{sec:method}

\begin{table*}[t]
\caption{Controls and the decomposition under both identifier schemes on
Beauty and Sports; three-seed means, the SFT start one checkpoint.
``top-1'': number of distinct top-1 items; Gini: Gini coefficient of
full-catalog exposure; cov.: coverage; NDCG and coverage at cutoff 10;
arrows give the direction of improvement. The last row of each block is
per-item centering (\S\ref{sub:baselines}).}
\label{tab:controls}
\centering
\small\setlength{\tabcolsep}{3pt}\renewcommand{\arraystretch}{0.82}
\resizebox{\textwidth}{!}{%
\begin{tabular}{llrrrrrrrrrr}
\toprule
& & \multicolumn{5}{c}{Beauty} & \multicolumn{5}{c}{Sports} \\
\cmidrule(lr){3-7}\cmidrule(lr){8-12}
identifiers & arm & HR@10$\uparrow$ & NDCG$\uparrow$ & top-1$\uparrow$ & Gini$\downarrow$ & cov.$\uparrow$ & HR@10$\uparrow$ & NDCG$\uparrow$ & top-1$\uparrow$ & Gini$\downarrow$ & cov.$\uparrow$ \\
\midrule
\multirow{6}{*}{Semantic IDs} & SFT start & 0.0851 & 0.0491 & 1317 & 0.916 & 0.436 & 0.0473 & 0.0248 & 840 & 0.960 & 0.269 \\
 & random reward & 0.0849 & 0.0490 & 1313 & 0.916 & 0.435 & 0.0476 & 0.0250 & 840 & 0.961 & 0.268 \\
 & rule reward & 0.0876 & 0.0515 & 1659 & 0.881 & 0.537 & 0.0495 & 0.0264 & 788 & 0.955 & 0.304 \\
 & CF reward $r$ & 0.0843 & 0.0473 & 837 & 0.947 & 0.404 & 0.0418 & 0.0202 & 169 & 0.990 & 0.152 \\
 & $\hat m$ only & 0.0318 & 0.0150 & 32 & 0.998 & 0.061 & 0.0306 & 0.0146 & 15 & 0.997 & 0.085 \\
 & per-item centering ($r-\hat m$) & 0.0813 & 0.0465 & 1830 & 0.879 & 0.577 & 0.0433 & 0.0211 & 699 & 0.971 & 0.270 \\
\midrule
\multirow{6}{*}{Atomic IDs} & SFT start & 0.0899 & 0.0487 & 857 & 0.940 & 0.325 & 0.0487 & 0.0260 & 463 & 0.975 & 0.163 \\
 & random reward & 0.0902 & 0.0490 & 878 & 0.939 & 0.328 & 0.0485 & 0.0260 & 459 & 0.975 & 0.160 \\
 & rule reward & 0.0925 & 0.0501 & 953 & 0.933 & 0.353 & 0.0501 & 0.0267 & 406 & 0.978 & 0.149 \\
 & CF reward $r$ & 0.0888 & 0.0483 & 544 & 0.962 & 0.237 & 0.0496 & 0.0260 & 168 & 0.990 & 0.083 \\
 & $\hat m$ only & 0.0816 & 0.0446 & 346 & 0.978 & 0.171 & 0.0440 & 0.0224 & 130 & 0.993 & 0.069 \\
 & per-item centering ($r-\hat m$) & 0.0904 & 0.0494 & 829 & 0.943 & 0.316 & 0.0506 & 0.0270 & 266 & 0.985 & 0.108 \\
\bottomrule
\end{tabular}}

\end{table*}

Reward decomposition links exposure collapse to shared scoring
tendencies in CF rewards. This motivates two corrections: per-item
centering removes the item mean, and per-item standardization further
adjusts the residual score scale.

\subsection{CF Reward Decomposition}
\label{sub:decomp}

Let $u$ denote a user context represented by its interaction history,
and $Q$ the uniform empirical distribution over training examples with
valid, nonempty histories. Using the same $Q$ for all items, the CF
score is decomposed as
\begin{equation}
\begin{aligned}
r(u,i)&=m(i)+\varepsilon(u,i),\\
m(i)&=\mathbb E_{u\sim Q}[r(u,i)].
\end{aligned}
\label{eq:decomp}
\end{equation}
The \textbf{\textit{item-mean component}} $m(i)$ captures the CF model's
average scoring tendency toward item $i$ across the reference histories.
The \textbf{\textit{residual component}}
$\varepsilon(u,i)=r(u,i)-m(i)$ captures the current user context's score
deviation from that average and satisfies
$\mathbb E_{u\sim Q}[\varepsilon(u,i)]=0$.

Controlled experiments show that training on the estimated item-mean
component alone produces exposure collapse, whereas training on the
estimated residual component broadens exposure relative to the
uncorrected CF reward
(\S\ref{sub:component}). These findings identify the item-mean component
as a contributor to exposure concentration and motivate its removal
from the training reward.

\subsection{Per-Item Centering}
\label{sub:baselines}

\begin{figure}[t]
\centering
\includegraphics[width=\columnwidth]{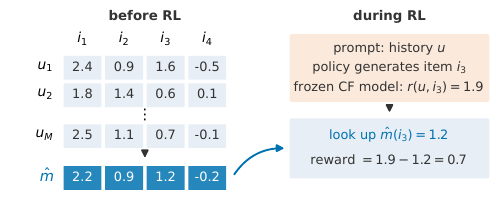}
\caption{Per-item centering: before RL, the frozen CF model scores the
catalog for $M$ sampled training histories and the scores are averaged
per item into $\hat m$; during RL, the reward for a generated item is its
CF score minus $\hat m$. Illustrative values.}
\label{fig:centering}
\end{figure}

For finite scores on the reference set, the item mean is the
least-squares projection of an item's scores onto
constants~\cite{books/daglib/0073491}:
\begin{equation}
m(i)=\arg\min_{c\in\mathbb R}
\mathbb E_{u\sim Q}\!\left[(r(u,i)-c)^2\right].
\label{eq:item_projection}
\end{equation}
Before RL, $M\geq1$ training-history examples $u_1,\ldots,u_M$ are
sampled uniformly without replacement from the reference pool. The CF
model scores the full catalog for each history
(Figure~\ref{fig:centering}), yielding the estimated item mean and the
\textbf{\textit{per-item centering}} reward:
\begin{equation}
\begin{aligned}
\hat m(i)&=\frac1M\sum_{j=1}^{M}r(u_j,i),\\
r_{\mathrm{centered}}(u,i)&=r(u,i)-\hat m(i).
\end{aligned}
\label{eq:center}
\end{equation}
This reward has zero mean over the sampled histories and mean
$m(i)-\hat m(i)$ under $Q$. Because the subtraction depends on the
generated item, it can change reward differences between items and
constitutes a reward intervention~\cite{tucker2018mirageactiondependentbaselinesreinforcement}.

\subsection{Per-Item Standardization}
\label{sub:std}

Centering leaves each item's score variance across user histories
unchanged, so residual components can still differ in scale. Following
z-score normalization in neighborhood-based
CF~\cite{herlocker1999cf,nikolakopoulos2021trustneighborscomprehensivesurvey},
\textbf{\textit{per-item standardization}} further corrects these scale
differences using the same sampled histories:
\begin{equation}
\begin{aligned}
\hat\sigma^2(i)&=\frac1M\sum_{j=1}^{M}
\bigl(r(u_j,i)-\hat m(i)\bigr)^2,\\
r_{\mathrm{std}}(u,i)&=
\frac{r(u,i)-\hat m(i)}{\max(\hat\sigma(i),s_0)},
\qquad s_0>0.
\end{aligned}
\label{eq:standardized}
\end{equation}
The floor $s_0$ limits amplification when an item's estimated standard
deviation is small. The main experiments use the tenth percentile of
$\hat\sigma$ over the catalog, which is positive in all four settings.

Let $\hat Q$ denote the uniform empirical distribution over the sampled
histories. For each item, the corrected reward satisfies
\begin{equation}
\begin{aligned}
\mathbb E_{\hat Q}[r_{\mathrm{std}}(u,i)]&=0,\\
\operatorname{Var}_{\hat Q}[r_{\mathrm{std}}(u,i)]
&=\frac{\hat\sigma^2(i)}
{\max(\hat\sigma(i),s_0)^2}\leq 1.
\end{aligned}
\label{eq:empirical_standardization}
\end{equation}
Standardization therefore preserves empirical centering while
controlling the residual score scale. We evaluate its effect on exposure
by comparing with centering and varying the floor
(\S\ref{sub:family_exp}).

Both corrections use item statistics computed once before RL and held
fixed alongside the CF model. During training, the stored statistics
transform CF scores into rewards, with the same policy optimizer and
training budget as the uncorrected CF arm.

\section{Experiments}
\label{sec:experiments}

The experiments examine exposure collapse under CF-reward training,
identify the role of the item-mean component through controlled interventions,
and evaluate the effectiveness and generality of per-item reward corrections.

\subsection{Experimental Setup}

\paragraph{\textbf{Datasets and protocol.}}
Experiments use Amazon Beauty~\cite{mcauley2015imagebasedrecommendationsstylessubstitutes}, with 15{,}750 test
users and 7{,}819 items, and Amazon Sports and Outdoors, with 19{,}679 test
users and 10{,}052 items. Each user's interaction sequence follows a
leave-one-out split: the last item is the test target, the second-to-last
is the validation target, and earlier prefixes form the training examples.
The policy receives up to 20 preceding items and generates the next item's
identifier. Evaluation uses the full catalog without filtering previously
interacted items. A MiniOneRec-based training
pipeline~\cite{kong2025minionerecopensourceframeworkscaling} supports the two evaluated identifier
schemes, Semantic IDs and Atomic IDs, yielding four main settings.

\paragraph{\textbf{Models and reward controls.}}
The backbone is Qwen2.5-0.5B-Instruct~\cite{qwen2025qwen25technicalreport}.
SASRec~\cite{kang2018selfattentivesequentialrecommendation} is trained on the training windows and frozen
before RL; its raw logit for a generated item, conditioned on the user's
interaction history, supplies the CF reward. CF rewards are available in
MiniOneRec's implementation and evaluated in its reward ablation. Two
additional rewards provide controls: the \textbf{\textit{rule reward}}
uses the exact-match and rank-aware reward combination selected by
MiniOneRec's released RL script, while the \textbf{\textit{random reward}}
assigns an independent $\mathcal N(0,1)$ draw to each valid generation.
A policy trained under a given reward is termed an
\textbf{\textit{arm}}. Section~\ref{sub:scope} tests more CF architectures and LLM backbones.

\paragraph{\textbf{Comparison methods.}}
A stronger KL penalty~\cite{gao2022scalinglawsrewardmodel} constrains policy updates, whereas
Dr.\ GRPO~\cite{liu2025understandingr1zeroliketrainingcritical} changes the advantage estimator.
The HiLaR-style arm~\cite{hu2026hierarchicallatentreasoningllmbased} applies only its CF-reward
normalization: for each user, catalog scores are standardized and
clipped to $[-3,3]$. A popularity penalty~\cite{petrov2024aligninggptrecbeyondaccuracygoals} subtracts
$\lambda$ times the item's training-popularity quantile from its CF score.
The main comparison uses $\lambda=3$; Tables~\ref{tab:full}
and~\ref{tab:full_tok} also report results for $\lambda=1$.

\paragraph{\textbf{Training configuration.}}
Arms within each setting share the same SFT checkpoint. Unless otherwise
stated, RL uses GRPO with a group size of 16, 400 training steps, and a KL
coefficient of $10^{-3}$. The learning rate is $3\times10^{-6}$ for Atomic
IDs. For Semantic IDs, the rule arm uses $10^{-6}$, following the reference
configuration, and the other arms use $3\times10^{-7}$. Per-item statistics
are estimated from 4{,}000 sampled training histories and remain fixed
during RL. Appendix~\ref{app:impl} provides the identifier construction,
constrained decoding, task mixture, and implementation details.

\paragraph{\textbf{Evaluation.}}
Exposure is measured by distinct top-1 items, coverage@10, and full-catalog
exposure Gini, as defined in Section~\ref{sub:metrics}. Recommendation
accuracy is measured by HR@10 and NDCG@10. All metrics use the full test
split. Main-text tables report means over three RL seeds; the SFT start
is one checkpoint per setting, evaluated once. Figure~\ref{fig:breadth}
additionally reports standard deviations across seeds. Item popularity
is measured by the catalog quantile of its training interaction
count~\cite{Klimashevskaia_2024}; mid-tier targets have quantiles in
$[0.2,0.8)$. Accuracy comparisons pair the same test users within each seed,
using exact McNemar tests for hits and paired bootstrap tests for NDCG@10.
Appendix~\ref{app:tables} reports the paired comparisons.

\begin{table*}[t]
\centering
\begin{minipage}{\textwidth}
\caption{CF-reward corrections under the common base recipe with the
listed modifications; three-seed means. Metrics as in
Table~\ref{tab:controls}; bold: best in each column within each scheme,
including ties. HiLaR: only its per-user standardization and clipping.}
\label{tab:fix}
\centering
\begingroup
\small\setlength{\tabcolsep}{3pt}\renewcommand{\arraystretch}{0.82}
\resizebox{\textwidth}{!}{%
\begin{tabular}{llrrrrrrrrrr}
\toprule
& & \multicolumn{5}{c}{Beauty} & \multicolumn{5}{c}{Sports} \\
\cmidrule(lr){3-7}\cmidrule(lr){8-12}
identifiers & arm & HR@10$\uparrow$ & NDCG$\uparrow$ & top-1$\uparrow$ & Gini$\downarrow$ & cov.$\uparrow$ & HR@10$\uparrow$ & NDCG$\uparrow$ & top-1$\uparrow$ & Gini$\downarrow$ & cov.$\uparrow$ \\
\midrule
\multirow{7}{*}{Semantic IDs} & CF reward $r$ (GRPO) & 0.0843 & 0.0473 & 837 & 0.947 & 0.404 & 0.0418 & 0.0202 & 169 & 0.990 & 0.152 \\
 & Dr.\ GRPO~\cite{liu2025understandingr1zeroliketrainingcritical} & 0.0845 & 0.0477 & 956 & 0.941 & 0.425 & 0.0416 & 0.0202 & 155 & 0.990 & 0.151 \\
 & HiLaR std.~\cite{hu2026hierarchicallatentreasoningllmbased} & 0.0821 & 0.0461 & 699 & 0.955 & 0.358 & 0.0414 & 0.0201 & 205 & 0.990 & 0.148 \\
 & KL $\times 10$~\cite{gao2022scalinglawsrewardmodel} & 0.0843 & 0.0474 & 847 & 0.946 & 0.405 & 0.0426 & 0.0208 & 192 & 0.990 & 0.154 \\
 & pop.\ penalty $\lambda{=}3$~\cite{petrov2024aligninggptrecbeyondaccuracygoals} & \textbf{0.0852} & \textbf{0.0487} & 1377 & 0.908 & 0.515 & 0.0447 & 0.0219 & 349 & 0.979 & 0.228 \\
 & per-item centering ($r-\hat m$) & 0.0813 & 0.0465 & 1830 & 0.879 & 0.577 & 0.0433 & 0.0211 & 699 & 0.971 & 0.270 \\
 & per-item standardization ($r_{\mathrm{std}}$) & 0.0818 & 0.0460 & \textbf{1915} & \textbf{0.872} & \textbf{0.593} & \textbf{0.0464} & \textbf{0.0238} & \textbf{1652} & \textbf{0.930} & \textbf{0.436} \\
\midrule
\multirow{7}{*}{Atomic IDs} & CF reward $r$ (GRPO) & 0.0888 & 0.0483 & 544 & 0.962 & 0.237 & 0.0496 & 0.0260 & 168 & 0.990 & 0.083 \\
 & Dr.\ GRPO~\cite{liu2025understandingr1zeroliketrainingcritical} & 0.0887 & 0.0482 & 547 & 0.963 & 0.234 & 0.0497 & 0.0261 & 167 & 0.990 & 0.082 \\
 & HiLaR std.~\cite{hu2026hierarchicallatentreasoningllmbased} & 0.0884 & 0.0480 & 524 & 0.964 & 0.230 & 0.0492 & 0.0258 & 172 & 0.990 & 0.080 \\
 & KL $\times 10$~\cite{gao2022scalinglawsrewardmodel} & 0.0887 & 0.0483 & 543 & 0.963 & 0.236 & 0.0497 & 0.0261 & 169 & 0.990 & 0.082 \\
 & pop.\ penalty $\lambda{=}3$~\cite{petrov2024aligninggptrecbeyondaccuracygoals} & 0.0902 & 0.0490 & 729 & 0.950 & 0.296 & 0.0504 & 0.0266 & 214 & 0.988 & 0.098 \\
 & per-item centering ($r-\hat m$) & \textbf{0.0904} & \textbf{0.0494} & 829 & 0.943 & 0.316 & 0.0506 & 0.0270 & 266 & 0.985 & 0.108 \\
 & per-item standardization ($r_{\mathrm{std}}$) & 0.0903 & 0.0493 & \textbf{874} & \textbf{0.940} & \textbf{0.326} & \textbf{0.0513} & \textbf{0.0273} & \textbf{366} & \textbf{0.980} & \textbf{0.130} \\
\bottomrule
\end{tabular}}

\endgroup
\end{minipage}
\par\vspace{4pt}
\begin{minipage}{\textwidth}
\captionsetup*{type=figure}
\centering
\begingroup
\setlength{\tabcolsep}{0pt}
\renewcommand{\arraystretch}{1}
\begin{tabular}{@{}c@{\hspace{0.012\textwidth}}c@{\hspace{0.02\textwidth}}c@{}}
& {\small Beauty} & {\small Sports} \\[0.2ex]
\begin{minipage}[c]{0.08\textwidth}\centering\small Semantic\\IDs\end{minipage}
& \begin{minipage}[c]{0.32\textwidth}\includegraphics[width=\linewidth]{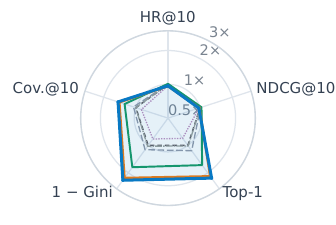}\end{minipage}
& \begin{minipage}[c]{0.32\textwidth}\includegraphics[width=\linewidth]{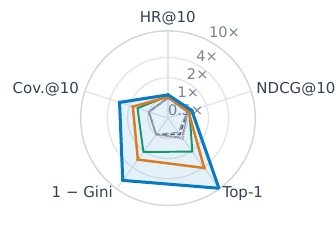}\end{minipage} \\[0.35ex]
\begin{minipage}[c]{0.08\textwidth}\centering\small Atomic\\IDs\end{minipage}
& \begin{minipage}[c]{0.32\textwidth}\includegraphics[width=\linewidth]{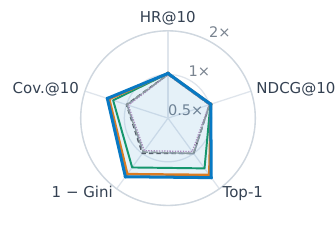}\end{minipage}
& \begin{minipage}[c]{0.32\textwidth}\includegraphics[width=\linewidth]{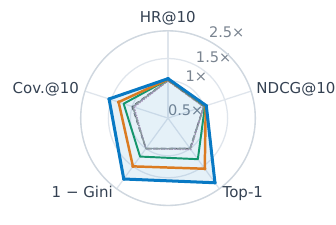}\end{minipage}
\end{tabular}
\par\vspace{0.4ex}
\includegraphics[width=0.86\textwidth]{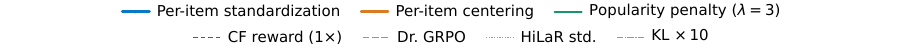}
\endgroup

\caption{Per-item standardization ranks first and per-item
centering second on all three exposure metrics among the methods in
Table~\ref{tab:fix} in all four settings. Axes show ratios of three-seed
means to the CF arm in the same setting, with Gini transformed to
$1-\mathrm{Gini}$; logarithmic radial scales are set independently for
each panel.}
\label{fig:fix}
\end{minipage}
\end{table*}

\subsection[RQ1: Does Training on CF Scores Concentrate Exposure?]{RQ1: Does Training on CF\\Scores Concentrate Exposure?}
\label{sub:problem}

\textbf{CF-reward training produces exposure collapse across all four main settings.}
Relative to the SFT start, the CF arm produces fewer distinct top-1 items,
higher Gini, and lower coverage@10 (Figure~\ref{fig:problem} and
Table~\ref{tab:controls}). With Semantic IDs, distinct top-1 items decrease
from 1{,}317 to 837 on Beauty and from 840 to 169 on Sports. The reductions
in all four settings meet the exposure-collapse criterion in
Section~\ref{sub:metrics}.

This contraction can coexist with small changes in aggregate accuracy.
On Beauty with Semantic IDs, HR@10 changes from 0.0851 to 0.0843, with no
statistically significant difference on any seed ($p\geq0.33$), despite a
net reduction of approximately 480 in the number of distinct top-1 items.
For users with held-out targets in the mid tier, HR@10 decreases
by approximately 9--56\% across the four settings, with a decrease on every
seed ($p<0.05$; Table~\ref{tab:tiers}).

In the seed-42 runs, items losing top-1 exposure are less popular than those
retaining it in all four settings. On Beauty with Semantic IDs, the
respective median popularity quantiles are 0.81 and 0.90.

\subsection[RQ2: What Do the Reward and Scorer Controls Reveal?]{RQ2: What Do the Reward and\\Scorer Controls Reveal?}
\label{sub:nothacking}

\textbf{The frozen CF scorer retains personalized ranking information even
when the policy trained on its scores exhibits concentrated exposure.}
When SASRec directly ranks the catalog for the same test users, it produces
3{,}407 distinct top-1 items on Beauty and 1{,}356 on Sports
(Table~\ref{tab:alone}). Its HR@10 is 0.0695 and 0.0381, respectively,
compared with 0.0130 and 0.0193 when every user receives the same ranking
based on estimated item means. These rankings confirm that the scorer
retains useful user-dependent information.

The alternative-reward controls examine whether the contraction also occurs
under the tested random and rule rewards (Table~\ref{tab:controls}).
With the remaining configuration matched to the CF arm, random rewards
leave the distinct top-1 count within 21 items of the SFT start in every
setting, with similar HR@10. The rule reward, using its reference learning
rate, increases the count on Beauty and keeps it within 13\% of the SFT
start on Sports, while achieving HR@10 at or above the start in all four
settings. Neither alternative reproduces the collapse under CF-reward
training.

\begin{table*}[t]
\caption{Four CF model architectures under both identifier schemes on
Beauty and Sports: the policy trained on the CF reward $r$, on per-item
centering ($r-\hat m$) and on per-item standardization
($r_{\mathrm{std}}$); three-seed means; bold: the highest distinct top-1
count.}
\label{tab:scorers}
\centering
\small\setlength{\tabcolsep}{3.5pt}\renewcommand{\arraystretch}{0.82}
\resizebox{\textwidth}{!}{%
\begin{tabular}{llrrrrrrrrrrrr}
\toprule
& & \multicolumn{6}{c}{Beauty} & \multicolumn{6}{c}{Sports} \\
\cmidrule(lr){3-8}\cmidrule(lr){9-14}
& & \multicolumn{3}{c}{distinct top-1 items$\uparrow$} & \multicolumn{3}{c}{HR@10$\uparrow$} & \multicolumn{3}{c}{distinct top-1 items$\uparrow$} & \multicolumn{3}{c}{HR@10$\uparrow$} \\
\cmidrule(lr){3-5}\cmidrule(lr){6-8}\cmidrule(lr){9-11}\cmidrule(lr){12-14}
identifiers & CF model & $r$ & $r-\hat m$ & $r_{\mathrm{std}}$ & $r$ & $r-\hat m$ & $r_{\mathrm{std}}$ & $r$ & $r-\hat m$ & $r_{\mathrm{std}}$ & $r$ & $r-\hat m$ & $r_{\mathrm{std}}$ \\
\midrule
\multirow{4}{*}{Semantic IDs} & SASRec & 837 & 1830 & \textbf{1915} & 0.0843 & 0.0813 & 0.0818 & 169 & 699 & \textbf{1652} & 0.0418 & 0.0433 & 0.0464 \\
 & GRU4Rec & 864 & 1635 & \textbf{1707} & 0.0864 & 0.0828 & 0.0825 & 207 & 593 & \textbf{1089} & 0.0434 & 0.0435 & 0.0467 \\
 & two-tower & 502 & 2184 & \textbf{2486} & 0.0769 & 0.0752 & 0.0759 & 412 & 1579 & \textbf{3079} & 0.0475 & 0.0453 & 0.0392 \\
 & CTR head & 281 & 1540 & \textbf{1804} & 0.0702 & 0.0717 & 0.0766 & 43 & 737 & \textbf{1095} & 0.0352 & 0.0350 & 0.0429 \\
\midrule
\multirow{4}{*}{Atomic IDs} & SASRec & 544 & 829 & \textbf{874} & 0.0888 & 0.0904 & 0.0903 & 168 & 266 & \textbf{366} & 0.0496 & 0.0506 & 0.0513 \\
 & GRU4Rec & 553 & 817 & \textbf{887} & 0.0886 & 0.0906 & 0.0905 & 180 & 281 & \textbf{373} & 0.0507 & 0.0514 & 0.0519 \\
 & two-tower & 503 & \textbf{1019} & 889 & 0.0877 & 0.0906 & 0.0896 & 183 & 263 & \textbf{384} & 0.0504 & 0.0509 & 0.0505 \\
 & CTR head & 454 & \textbf{1025} & 911 & 0.0859 & 0.0890 & 0.0893 & 134 & 433 & \textbf{458} & 0.0474 & 0.0479 & 0.0489 \\
\bottomrule
\end{tabular}}

\end{table*}

\subsection[RQ3: How Does the Item-Mean Component Affect Exposure?]{RQ3: How Does the Item-Mean\\Component Affect Exposure?}
\label{sub:component}

\textbf{Controlled interventions identify the item-mean component as a
contributor to exposure concentration.}
The interventions train separately on the estimated item-mean component
$\hat m(i)$ and the estimated residual component $r(u,i)-\hat m(i)$,
with the remaining
training settings fixed (Table~\ref{tab:controls}). Training on the
estimated item-mean component alone produces collapse in all four
settings, leaving only 32 distinct top-1 items on Beauty and 15 on Sports
with Semantic IDs. Training on the estimated residual component, equivalent
to per-item centering, broadens exposure relative to the CF arm in every
setting; on Beauty with Semantic IDs, the count increases from 837 to
1{,}830. These separate-component interventions do not quantify each
component's share of the original collapse.

The estimated means of top-ranked items provide supporting evidence.
For each user, Table~\ref{tab:spread} records the catalog percentile of the
estimated mean of the top-ranked item and averages these percentiles
across users. This statistic increases under CF-reward training in all
four settings: with Semantic IDs, it rises from 0.84 to 0.90 on Beauty and
from 0.91 to 0.97 on Sports. Both per-item corrections reduce it relative
to the CF arm.

\subsection{RQ4: How Effectively Do Per-Item Corrections Broaden Exposure?}
\label{sub:fix}

\textbf{Both per-item corrections improve all three exposure metrics relative
to the uncorrected CF reward across the four main settings.}
Standardization produces 1.6--9.8 times as many distinct top-1 items as the
CF arm (Table~\ref{tab:fix} and Figure~\ref{fig:fix}).
On Sports with Semantic IDs, the count increases
from 169 under the CF reward to 699 with centering and 1{,}652 with
standardization. Among the configurations in Table~\ref{tab:fix},
standardization has the most distinct top-1 items, the highest coverage@10,
and the lowest Gini, followed by centering, in every setting.

Increasing the KL coefficient from $10^{-3}$ to $10^{-2}$ changes the
distinct top-1 count by no more than about 23 items. Dr.\ GRPO and
HiLaR-style standardization do not consistently increase this count with
Semantic IDs; with Atomic IDs, both remain within 20 items of the CF arm.

At $\lambda=3$, the popularity penalty broadens exposure, but both per-item
corrections produce more distinct top-1 items in all four settings.
Tables~\ref{tab:full} and~\ref{tab:full_tok} report accuracy and exposure
for $\lambda \in \{1,3\}$. Across all four settings, per-item corrections
share the mean-removal rule, and standardization uses the same floor
quantile.

Recommendation accuracy varies across settings, with full paired
comparisons provided in Appendix~\ref{app:tables}. For mid-tier targets,
standardization raises HR@10 to approximately 1.3--2.8 times that of the CF
arm, with improvements in all four settings and on every seed
($p<10^{-5}$; Table~\ref{tab:tiers}).

\subsection[RQ5: How Does the Correction Design Affect Exposure?]{RQ5: How Does the Correction\\Design Affect Exposure?}
\label{sub:family_exp}
\label{sub:family}

\begin{figure*}[t]
\centering
\includegraphics[width=\textwidth]{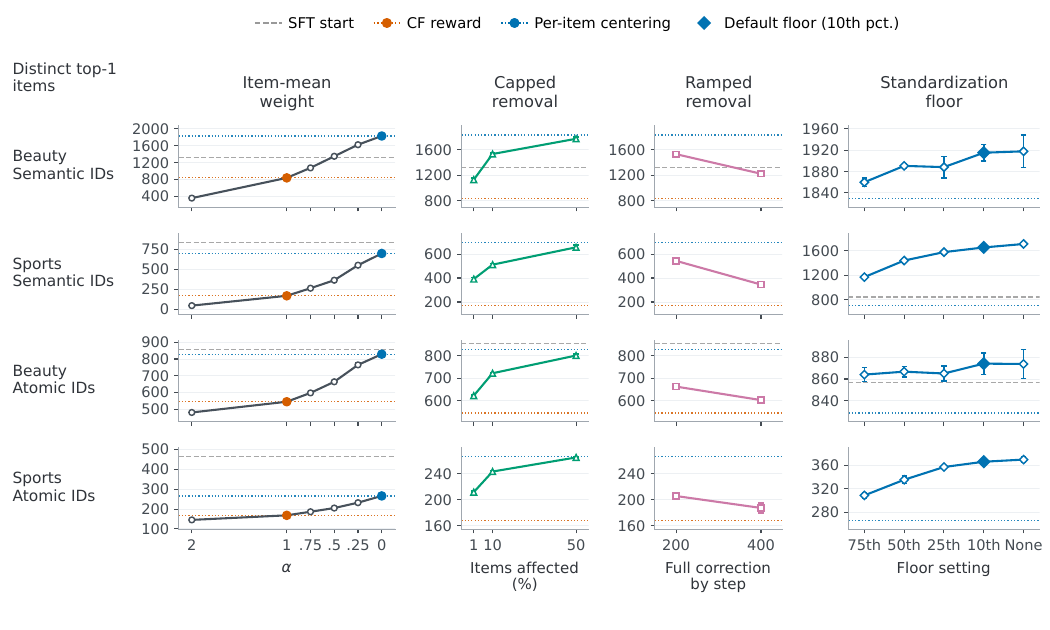}
\caption{Distinct top-1 items after 400 steps (three-seed mean $\pm$ SD).
Rows: four main settings; columns: four reward probes with independent
vertical ranges. Ramp labels: full-correction steps; floor labels:
$\hat\sigma$ percentiles (\texttt{None}: no floor).
Every tested floor exceeds centering (\S\ref{sub:family};
full results in Table~\ref{tab:breadth}).}
\label{fig:breadth}
\end{figure*}

\textbf{Across the tested item-mean weights, removing more of the estimated
item-mean component consistently broadens exposure.}
Four interventions examine correction strength, which items are corrected,
correction timing, and the standardization floor (Figure~\ref{fig:breadth}
and Table~\ref{tab:breadth}). They reuse the same estimated item statistics
and training configuration. For the capped and ramped variants, the
recovery fraction is the gain in distinct top-1 items over the CF arm
divided by the gap between centering and the CF arm.

Item-mean weighting varies the correction strength through
\[
r_{\alpha}(u,i)=r(u,i)-(1-\alpha)\hat m(i),
\]
where $\alpha=1$ gives the uncorrected CF reward, $\alpha=0$ gives per-item
centering, and $\alpha=2$ doubles the estimated item-mean contribution.
As $\alpha$ decreases through the tested values from 2 to 0, distinct top-1
items increase and Gini decreases at every step in all four settings.
On Beauty with Semantic IDs, the respective counts at
$\alpha=2,1,0$ are 358, 837, and 1{,}830.

Capped removal restricts the correction to high-mean items:
\[
r_{\mathrm{cap}}(u,i)
=r(u,i)-\max\{\hat m(i)-\tau_q,0\},
\]
where $\tau_q$ is a catalog quantile of $\hat m$. Only items above this
threshold are corrected. Targeting the top 1\%, 10\%, and 50\% of items
by estimated mean recovers approximately 27--44\%, 63--77\%, and 91--99\%
of the gap between the CF arm and per-item centering, respectively.

Ramped removal varies the correction over training:
\[
\begin{aligned}
r_t(u,i)&=r(u,i)-\beta_t\hat m(i),\\
\beta_t&=\min(1,t/T),
\end{aligned}
\]
with $T=200$ or $400$ in a 400-step run. The 200-step ramp recovers 70--71\%
of the gap with Semantic IDs and 38--42\% with Atomic IDs; the 400-step ramp
recovers 19--39\% across the four settings. Both ramps produce fewer
distinct top-1 items than constant partial removal with approximately
matched average strength, using $\alpha=0.25$ and $\alpha=0.5$,
respectively.

For the scale correction, the floor $s_0$ is varied over the 10th, 25th,
50th, and 75th catalog percentiles of $\hat\sigma$, together with a
no-floor configuration.
All five configurations produce more distinct top-1 items than centering
in every setting, although the size of the gain varies.
On Sports with Semantic IDs, they yield
1{,}167--1{,}709 distinct top-1 items, compared with 699 for centering.
Within each setting, the range of mean HR@10 across these configurations
is at most 0.0008. On Sports, standardization also lowers the average
catalog percentile of the estimated score standard deviation for users'
top-ranked items relative to centering (Table~\ref{tab:spread}).

\subsection{RQ6: Do the Findings Generalize across CF Models and LLM Backbones?}
\label{sub:scope}

\textbf{Both corrections increase distinct top-1 items across the evaluated
CF architectures and LLM backbones.}
Table~\ref{tab:scorers} evaluates SASRec, GRU4Rec, a two-tower model, and a
pointwise CTR model across the four main settings. Training on their
uncorrected scores reduces distinct top-1 items relative to the SFT start
in all 16 combinations. Both per-item corrections increase the count
relative to the corresponding CF arm in every combination, and
standardization exceeds centering in 14 of the 16. Centering's HR@10
remains within 0.004 of the corresponding CF arm throughout.

The additional-backbone experiments use Beauty with Semantic IDs.
CF-reward training reduces distinct top-1
items by about 23--41\% for Qwen2.5-1.5B-Instruct and
Llama-3.2-1B-Instruct, even as HR@10 increases. Both corrections broaden
exposure relative to the CF arm, with HR@10 within 0.002 of it
(Table~\ref{tab:backbones}).

\balance
\section{Conclusion}

Training LLM recommenders on CF scores can substantially concentrate
exposure even when the CF model retains useful personalized ranking
information. Reward decomposition and controlled interventions identify
the item-mean component as a contributor to this concentration.
Per-item centering subtracts its estimate, while per-item standardization
additionally normalizes the residual score scale. Across two datasets and
two identifier schemes, both corrections broaden exposure, increase
catalog coverage, and improve exposure evenness relative to the
uncorrected CF reward, without retraining the CF model. Further
evaluations across CF architectures and with two additional LLM
backbones on Beauty with Semantic IDs support the gains in distinct
top-1 items.

Evaluating CF rewards therefore requires examining the resulting
policy's exposure alongside the scorer's direct-ranking performance.
The evidence comes from the evaluated critic-free RL configurations
with frozen CF scorers; recommendation accuracy varies across settings.
Future work could develop adaptive reward corrections for evolving user
and item distributions and investigate their effects on long-term
exposure and recommendation quality.

\section{Ethical Considerations}

Experiments use public Amazon review datasets and are conducted offline.
Reuse of interaction histories should respect dataset terms and user privacy.
Broader catalog exposure alone does not establish fairness across users or
item providers~\cite{singh2018fairness}. Deployment requires evaluating recommendation quality and
stakeholder impacts beyond the exposure metrics studied here.

\newpage
\nobalance

\balance
\bibliographystyle{ACM-Reference-Format}
\begingroup
\catcode`\&=12
\makeatletter
\LetLtxMacro{\originalshoweprint}{\showeprint}
\renewcommand{\showeprint}[2][arxiv]{%
  \begingroup
  \in@{https://doi.org/}{#2}%
  \ifin@
    \url{#2}%
  \else
    \originalshoweprint[#1]{#2}%
  \fi
  \endgroup}
\makeatother
\bibliography{refs}
\endgroup

\renewcommand{\dbltopfraction}{1.0}
\renewcommand{\dblfloatpagefraction}{0.8}
\setlength{\dblfloatsep}{6pt plus 2pt}
\setlength{\dbltextfloatsep}{8pt plus 2pt}
\clearpage
\nobalance
\appendix
\raggedbottom
\captionsetup[table]{skip=3pt,singlelinecheck=false}
\setlength{\intextsep}{4pt plus 1pt minus 1pt}
\setlength{\multicolsep}{6pt plus 2pt minus 1pt}

\section{Implementation Details}
\label{app:impl}

\begin{table}[H]
\caption{Distinct top-1 items with Atomic IDs at three RL learning rates
(main text: $3\times10^{-6}$).}
\label{tab:tokladder}
\centering
\small
\setlength{\tabcolsep}{3pt}
\renewcommand{\arraystretch}{1.05}

\begin{tabular*}{\linewidth}{@{\extracolsep{\fill}}lrrrrrr@{}}
\toprule
& \multicolumn{3}{c}{Beauty} & \multicolumn{3}{c}{Sports} \\
\cmidrule(lr){2-4}\cmidrule(lr){5-7}
arm & $3\times10^{-7}$ & $3\times10^{-6}$ & $10^{-5}$ & $3\times10^{-7}$ & $3\times10^{-6}$ & $10^{-5}$ \\
\midrule
random reward & 858 & 878 & 793 & 460 & 459 & 356 \\
CF reward $r$ & 776 & 544 & 482 & 337 & 168 & 133 \\
centering ($r-\hat m$) & 841 & 829 & 873 & 385 & 266 & 241 \\
$\hat m$ only & 705 & 346 & 194 & 326 & 130 & 50 \\
\bottomrule
\end{tabular*}

\end{table}

\paragraph{\textbf{Item identifiers.}} For Semantic IDs, item text (title, categories,
brand, description) is embedded and quantized by an RQ-VAE into three
semantic levels of 256 codes each, with a fourth level resolving collisions
so that the mapping is injective; decoding uses a prefix-constrained beam of
width 20 over a packed trie of legal code sequences, so every generated
sequence is a catalog item. For Atomic IDs each item is one dedicated
token and a sequence has length one; the trie, the beam and every other
setting are unchanged. The
Atomic-ID SFT uses a learning rate of $3\times10^{-4}$.

\paragraph{\textbf{Supervised fine-tuning.}} Learning rate $1\times10^{-5}$, cosine
schedule, 20 warmup steps, global batch 128, evaluation and checkpointing
every 5\% of training, early stopping with patience 3 on validation loss,
best checkpoint loaded at the end.

\paragraph{\textbf{CF models.}} All four are trained on the same leave-one-out
training windows as the policy and frozen before RL: SASRec with 2
blocks, 2 heads, hidden size 64, length 50 and dropout 0.2; GRU4Rec with
hidden size 64 and length 50; a two-tower model with a mean-pooled
history embedding and an item embedding of size 64; and a pointwise CTR
head on history and item embeddings of size 64. The first three use
next-item cross-entropy over the catalog, the CTR head binary
cross-entropy with four uniform negatives per positive; all use Adam at
learning rate $10^{-3}$, batch 256, at most 100 epochs, early stopping
with patience 5 on validation HR@10, best checkpoint kept. Each scores a
generated item by its raw logit at the last position of the user's
history, unscaled. An unparseable generation
receives the format penalty, $-1$, in every CF-based arm. Prompts without
an item history, the title-to-identifier tasks of the reference RL mix
(16\% of groups on Beauty, 20\% on Sports), cannot be scored by a
sequential CF model and receive the exact-match reward in every CF-based
arm. The HiLaR arm
implements only that method's per-user standardization of its CF reward
term: each user's logits are standardized over the full catalog, clipped
at $\pm 3$, and read off at the generated item.  The popularity penalty
subtracts $\lambda$ times the item's popularity quantile in $[0,1]$ from
the reward.

\paragraph{\textbf{GRPO}} Group size 16, 400 steps, 64 prompts per step (32 per
device, two accumulation steps), KL coefficient $10^{-3}$ with the
low-variance estimator of GRPO~\cite{shao2024deepseekmathpushinglimitsmathematical}, clip ratio 0.2, and the standard advantage
$A(x, y_j) = (r(x, y_j) - \bar r_x)/(\sigma_x + \epsilon)$ with $\bar r_x$
and $\sigma_x$ the mean and standard deviation of the rewards of the $G$
completions of prompt $x$. Rollouts follow the reference implementation:
the $G = 16$ completions of a prompt are the beams of a sampled beam
search (\texttt{num\_beams} $= G$, sampling on, temperature 1) under the
same prefix trie as evaluation, so every rollout is a catalog item; the
checkpoint's default top-$k$, top-$p$ and repetition penalty are
overridden to none, and prompts are generated in chunks of 16. The final
checkpoint is evaluated once.
$\hat{m}$ is estimated from 4000 training histories sampled with the
run's seed and scored against the full catalog, one forward pass per
history for the sequential and two-tower models and one batched pass
per history--item pair for the CTR head.

\paragraph{\textbf{Estimator variants.}} With the reward itself as the advantage,
no group mean and no standard deviation, at the same group size and
budget, the CF arm leaves 637 and 239 distinct top-1 items on Beauty
and Sports with Semantic IDs and 616 and 175 with Atomic IDs; with group
size 2 at the same prompts per step, one eighth of the rollouts, it
leaves 696 and 152, and 720 and 170. Both lose HR@10 that the standard
advantage does not.

\paragraph{\textbf{Variants of the case study.}} The variants of \S\ref{sub:family}
reuse the per-item centering run's recipe, SFT start, CF model and
seed in each setting, and the same $M = 4000$ histories produce $\hat m$
and $\hat\sigma$. For capped removal, $\tau_q$ is the $q$-quantile of
$\hat m$ over the catalog; at $q = 0.99$ it touches 79 items on Beauty
and 101 on Sports and 18--29\% of the scored rollouts, 43--58\% at
$q = 0.90$. For ramped
removal, $\beta_t = \min(1, t/T)$ at step $t$ of 400, with $T = 200$ or
$400$. The floor $s_0$ at the tenth percentile of $\hat\sigma$ is 1.65 on
Beauty (median 2.14) and 0.74 on Sports (median 1.17). 
The standardization arms of Table~\ref{tab:scorers} and of the two
backbones were run after the main batch; the two backbone standardization
runs use activation checkpointing and a CPU-offloaded AdamW (DeepSpeed
ZeRO-2) with the same hyperparameters, a path that reproduces the original
per-step gradient norm and KL on the 0.5B model.

\section{Additional Results}
\label{app:more}

\begin{table}[H]
\caption{The four CF models ranking the full catalog directly for the
test users, the SASRec rows being the rankings used in
\S\ref{sub:nothacking}; metrics as in Table~\ref{tab:controls}.}
\label{tab:alone}
\centering
\small
\setlength{\tabcolsep}{3pt}
\renewcommand{\arraystretch}{1.05}

\begin{tabular*}{\linewidth}{@{\extracolsep{\fill}}llrrrrr@{}}
\toprule
dataset & CF model & HR@10$\uparrow$ & NDCG$\uparrow$ & top-1$\uparrow$ & Gini$\downarrow$ & cov.$\uparrow$ \\
\midrule
\multirow{4}{*}{Beauty} & SASRec & 0.0695 & 0.0359 & 3407 & 0.778 & 0.702 \\
 & GRU4Rec & 0.0613 & 0.0351 & 2469 & 0.807 & 0.579 \\
 & two-tower & 0.0670 & 0.0289 & 4728 & 0.709 & 0.935 \\
 & CTR head & 0.0370 & 0.0173 & 2322 & 0.850 & 0.669 \\
\midrule
\multirow{4}{*}{Sports} & SASRec & 0.0381 & 0.0189 & 1356 & 0.961 & 0.314 \\
 & GRU4Rec & 0.0321 & 0.0166 & 603 & 0.971 & 0.174 \\
 & two-tower & 0.0401 & 0.0182 & 2191 & 0.909 & 0.536 \\
 & CTR head & 0.0243 & 0.0121 & 793 & 0.963 & 0.342 \\
\bottomrule
\end{tabular*}

\end{table}

\begin{table}[H]
\caption{CF rewards raise first-ranked items' $\hat m$ and $\hat\sigma$
percentiles relative to SFT; both corrections lower them in every setting.
Three-seed means; catalog correlations.
SID: Semantic IDs; AID: Atomic IDs.}
\label{tab:spread}
\centering
\small
\setlength{\tabcolsep}{3pt}
\renewcommand{\arraystretch}{1.05}

\setlength{\tabcolsep}{2.5pt}
\begin{tabular*}{\linewidth}{@{\extracolsep{\fill}}lrrrrrrrrr@{}}
\toprule
& & \multicolumn{4}{c}{percentile of $\hat m$} & \multicolumn{4}{c}{percentile of $\hat\sigma$} \\
\cmidrule(lr){3-6}\cmidrule(lr){7-10}
setting & corr$(\hat m,\hat\sigma)$ & start & CF & cent. & std. & start & CF & cent. & std. \\
\midrule
Beauty / SID & -0.22 & 0.84 & 0.90 & 0.80 & 0.79 & 0.64 & 0.71 & 0.65 & 0.63 \\
Sports / SID & -0.56 & 0.91 & 0.97 & 0.90 & 0.88 & 0.75 & 0.91 & 0.86 & 0.71 \\
Beauty / AID & -0.22 & 0.84 & 0.89 & 0.84 & 0.84 & 0.68 & 0.75 & 0.70 & 0.69 \\
Sports / AID & -0.56 & 0.93 & 0.97 & 0.95 & 0.94 & 0.80 & 0.89 & 0.87 & 0.84 \\
\bottomrule
\end{tabular*}

\end{table}

\onecolumn
\captionsetup[table]{margin=0pt,justification=raggedright,singlelinecheck=false}
\begin{multicols}{2}
\begin{table}[H]
\caption{HR@10 of the users whose held-out item is in the mid tier,
popularity quantile 0.2 to 0.8 (6586 users on Beauty, 7563 on Sports),
beside each arm's distinct top-1 items over all users; three-seed means,
the SFT start one checkpoint.}
\label{tab:tiers}
\centering
\small
\setlength{\tabcolsep}{3pt}
\renewcommand{\arraystretch}{1.05}

\setlength{\tabcolsep}{2pt}
\begin{tabular*}{\linewidth}{@{\extracolsep{\fill}}lrrrrrrrr@{}}
\toprule
& \multicolumn{4}{c}{Semantic IDs} & \multicolumn{4}{c}{Atomic IDs} \\
\cmidrule(lr){2-5}\cmidrule(lr){6-9}
& \multicolumn{2}{c}{Beauty} & \multicolumn{2}{c}{Sports} & \multicolumn{2}{c}{Beauty} & \multicolumn{2}{c}{Sports} \\
\cmidrule(lr){2-3}\cmidrule(lr){4-5}\cmidrule(lr){6-7}\cmidrule(lr){8-9}
arm & HR@10 & top-1 & HR@10 & top-1 & HR@10 & top-1 & HR@10 & top-1 \\
\midrule
start & 0.0346 & 1317 & 0.0123 & 840 & 0.0396 & 857 & 0.0097 & 463 \\
CF $r$ & 0.0315 & 837 & 0.0054 & 169 & 0.0284 & 544 & 0.0051 & 168 \\
pen.\ $\lambda{=}3$ & 0.0364 & 1377 & 0.0100 & 349 & 0.0352 & 729 & 0.0060 & 214 \\
$r-\hat m$ & 0.0392 & 1830 & 0.0114 & 699 & 0.0373 & 829 & 0.0065 & 266 \\
$r_{\mathrm{std}}$ & 0.0401 & 1915 & 0.0149 & 1652 & 0.0388 & 874 & 0.0077 & 366 \\
\bottomrule
\end{tabular*}

\end{table}

\paragraph{\textbf{Reference sample size.}}
Using 500 or 16{,}000 histories instead of 4000 to estimate $\hat m$
gives 1823 and 1847 distinct top-1 items on Beauty with Semantic IDs,
respectively, compared with 1830 at 4000 histories; HR@10 differs by
at most 0.0006.

\paragraph{\textbf{Item mean and popularity.}} Over the catalog, the squared
Pearson correlation between log interaction count and SASRec's $\hat m$ is 0.38 on Beauty and 0.67 on Sports.

\paragraph{\textbf{The Atomic-ID arms at other rates.}} Table~\ref{tab:tokladder}
repeats four arms of Table~\ref{tab:controls} with Atomic IDs at a tenth
of the main-text rate (the CF arm's realized KL on Beauty is 0.47 there)
and at about three times it; at every rate per-item centering leaves more
distinct top-1 items than the CF arm, and $\hat m$ alone the fewest.

\end{multicols}

\section{Full Results}
\label{app:tables}

\begin{table}[H]
\caption{Full metric set with Semantic IDs: within each dataset block,
accuracy precedes exposure; NDCG and coverage use cutoff 10, and arrows
indicate the direction of improvement.}
\label{tab:full}
\centering
\small
\setlength{\tabcolsep}{3pt}
\renewcommand{\arraystretch}{1.05}

\begin{tabular*}{\linewidth}{@{\extracolsep{\fill}}lrrrrrrrrrr@{}}
\toprule
& \multicolumn{5}{c}{Beauty} & \multicolumn{5}{c}{Sports} \\
\cmidrule(lr){2-6}\cmidrule(lr){7-11}
arm & HR@10$\uparrow$ & NDCG$\uparrow$ & Gini$\downarrow$ & cov.$\uparrow$ & top-1$\uparrow$ & HR@10$\uparrow$ & NDCG$\uparrow$ & Gini$\downarrow$ & cov.$\uparrow$ & top-1$\uparrow$ \\
\midrule
SFT start & 0.0851 & 0.0491 & 0.916 & 0.436 & 1317 & 0.0473 & 0.0248 & 0.960 & 0.269 & 840 \\
random reward & 0.0849 & 0.0490 & 0.916 & 0.435 & 1313 & 0.0476 & 0.0250 & 0.961 & 0.268 & 840 \\
rule reward & 0.0876 & 0.0515 & 0.881 & 0.537 & 1659 & 0.0495 & 0.0264 & 0.955 & 0.304 & 788 \\
CF reward $r$ (GRPO) & 0.0843 & 0.0473 & 0.947 & 0.404 & 837 & 0.0418 & 0.0202 & 0.990 & 0.152 & 169 \\
$\hat m$ only & 0.0318 & 0.0150 & 0.998 & 0.061 & 32 & 0.0306 & 0.0146 & 0.997 & 0.085 & 15 \\
Dr.\ GRPO & 0.0845 & 0.0477 & 0.941 & 0.425 & 956 & 0.0416 & 0.0202 & 0.990 & 0.151 & 155 \\
HiLaR std. & 0.0821 & 0.0461 & 0.955 & 0.358 & 699 & 0.0414 & 0.0201 & 0.990 & 0.148 & 205 \\
KL $\times 10$ & 0.0843 & 0.0474 & 0.946 & 0.405 & 847 & 0.0426 & 0.0208 & 0.990 & 0.154 & 192 \\
pop.\ penalty $\lambda{=}1$ & 0.0855 & 0.0485 & 0.926 & 0.464 & 1124 & 0.0446 & 0.0217 & 0.984 & 0.197 & 276 \\
pop.\ penalty $\lambda{=}3$ & 0.0852 & 0.0487 & 0.908 & 0.515 & 1377 & 0.0447 & 0.0219 & 0.979 & 0.228 & 349 \\
\addlinespace[2pt]
per-item centering ($r-\hat m$) & 0.0813 & 0.0465 & 0.879 & 0.577 & 1830 & 0.0433 & 0.0211 & 0.971 & 0.270 & 699 \\
per-item standardization ($r_{\mathrm{std}}$) & 0.0818 & 0.0460 & 0.872 & 0.593 & 1915 & 0.0464 & 0.0238 & 0.930 & 0.436 & 1652 \\
\bottomrule
\end{tabular*}

\end{table}

\begin{table}[H]
\caption{Full metric set with Atomic IDs; layout as in
Table~\ref{tab:full}.}
\label{tab:full_tok}
\centering
\small
\setlength{\tabcolsep}{3pt}
\renewcommand{\arraystretch}{1.05}

\begin{tabular*}{\linewidth}{@{\extracolsep{\fill}}lrrrrrrrrrr@{}}
\toprule
& \multicolumn{5}{c}{Beauty} & \multicolumn{5}{c}{Sports} \\
\cmidrule(lr){2-6}\cmidrule(lr){7-11}
arm & HR@10$\uparrow$ & NDCG$\uparrow$ & Gini$\downarrow$ & cov.$\uparrow$ & top-1$\uparrow$ & HR@10$\uparrow$ & NDCG$\uparrow$ & Gini$\downarrow$ & cov.$\uparrow$ & top-1$\uparrow$ \\
\midrule
SFT start & 0.0899 & 0.0487 & 0.940 & 0.325 & 857 & 0.0487 & 0.0260 & 0.975 & 0.163 & 463 \\
random reward & 0.0902 & 0.0490 & 0.939 & 0.328 & 878 & 0.0485 & 0.0260 & 0.975 & 0.160 & 459 \\
rule reward & 0.0925 & 0.0501 & 0.933 & 0.353 & 953 & 0.0501 & 0.0267 & 0.978 & 0.149 & 406 \\
CF reward $r$ (GRPO) & 0.0888 & 0.0483 & 0.962 & 0.237 & 544 & 0.0496 & 0.0260 & 0.990 & 0.083 & 168 \\
$\hat m$ only & 0.0816 & 0.0446 & 0.978 & 0.171 & 346 & 0.0440 & 0.0224 & 0.993 & 0.069 & 130 \\
Dr.\ GRPO & 0.0887 & 0.0482 & 0.963 & 0.234 & 547 & 0.0497 & 0.0261 & 0.990 & 0.082 & 167 \\
HiLaR std. & 0.0884 & 0.0480 & 0.964 & 0.230 & 524 & 0.0492 & 0.0258 & 0.990 & 0.080 & 172 \\
KL $\times 10$ & 0.0887 & 0.0483 & 0.963 & 0.236 & 543 & 0.0497 & 0.0261 & 0.990 & 0.082 & 169 \\
pop.\ penalty $\lambda{=}1$ & 0.0896 & 0.0485 & 0.958 & 0.262 & 630 & 0.0500 & 0.0263 & 0.988 & 0.094 & 194 \\
pop.\ penalty $\lambda{=}3$ & 0.0902 & 0.0490 & 0.950 & 0.296 & 729 & 0.0504 & 0.0266 & 0.988 & 0.098 & 214 \\
\addlinespace[2pt]
per-item centering ($r-\hat m$) & 0.0904 & 0.0494 & 0.943 & 0.316 & 829 & 0.0506 & 0.0270 & 0.985 & 0.108 & 266 \\
per-item standardization ($r_{\mathrm{std}}$) & 0.0903 & 0.0493 & 0.940 & 0.326 & 874 & 0.0513 & 0.0273 & 0.980 & 0.130 & 366 \\
\bottomrule
\end{tabular*}

\end{table}

\clearpage
\begin{table}[H]
\caption{The two further backbones of \S\ref{sub:scope} on Beauty with
Semantic IDs: RL training on CF scores reduces distinct top-1 items,
and both per-item corrections raise the count above the SFT start;
metric order as in Table~\ref{tab:full}.}
\label{tab:backbones}
\centering
\small
\setlength{\tabcolsep}{3pt}
\renewcommand{\arraystretch}{1.05}

\begin{tabular*}{\linewidth}{@{\extracolsep{\fill}}lrrrrrrrrrr@{}}
\toprule
& \multicolumn{5}{c}{Qwen2.5-1.5B} & \multicolumn{5}{c}{Llama-3.2-1B} \\
\cmidrule(lr){2-6}\cmidrule(lr){7-11}
arm & HR@10$\uparrow$ & NDCG$\uparrow$ & Gini$\downarrow$ & cov.$\uparrow$ & top-1$\uparrow$ & HR@10$\uparrow$ & NDCG$\uparrow$ & Gini$\downarrow$ & cov.$\uparrow$ & top-1$\uparrow$ \\
\midrule
SFT start & 0.0744 & 0.0412 & 0.946 & 0.292 & 841 & 0.0811 & 0.0459 & 0.934 & 0.360 & 955 \\
CF reward $r$ & 0.0777 & 0.0424 & 0.952 & 0.338 & 648 & 0.0821 & 0.0450 & 0.957 & 0.321 & 560 \\
$r-\hat m$ & 0.0761 & 0.0419 & 0.915 & 0.448 & 1308 & 0.0829 & 0.0470 & 0.915 & 0.424 & 1112 \\
$r_{\mathrm{std}}$ & 0.0771 & 0.0421 & 0.909 & 0.470 & 1452 & 0.0829 & 0.0468 & 0.912 & 0.430 & 1173 \\
\bottomrule
\end{tabular*}

\end{table}

\begin{table}[H]
\caption{Correction probes (\S\ref{sub:family_exp}). Three-seed means;
HR@10 and top-1 include seed standard deviations.
$\alpha=1$: CF reward; $\alpha=0$: centering.
The main standardization floor is the tenth percentile.}
\label{tab:breadth}
\centering
\small
\setlength{\tabcolsep}{3pt}
\renewcommand{\arraystretch}{1.05}

\begin{tabular*}{\linewidth}{@{\extracolsep{\fill}}llrrrrrrrrrr@{}}
\toprule
 & & \multicolumn{5}{c}{Beauty} & \multicolumn{5}{c}{Sports} \\
\cmidrule(lr){3-7}\cmidrule(lr){8-12}
scheme & arm & HR@10$\uparrow$ & NDCG$\uparrow$ & top-1$\uparrow$ & Gini$\downarrow$ & cov.$\uparrow$ & HR@10$\uparrow$ & NDCG$\uparrow$ & top-1$\uparrow$ & Gini$\downarrow$ & cov.$\uparrow$ \\
\midrule
\multirow{14}{*}{Semantic IDs} & $\alpha=2$ (mean doubled) & 0.0737 $\pm$ 0.0005 & 0.0404 & 358 $\pm$ 7 & 0.978 & 0.262 & 0.0357 $\pm$ 0.0004 & 0.0168 & 46 $\pm$ 1 & 0.995 & 0.108 \\
 & $\alpha=0.75$ & 0.0848 $\pm$ 0.0005 & 0.0484 & 1075 $\pm$ 23 & 0.932 & 0.449 & 0.0443 $\pm$ 0.0005 & 0.0219 & 264 $\pm$ 4 & 0.987 & 0.171 \\
 & $\alpha=0.5$ & 0.0850 $\pm$ 0.0001 & 0.0487 & 1349 $\pm$ 39 & 0.915 & 0.495 & 0.0456 $\pm$ 0.0001 & 0.0233 & 364 $\pm$ 12 & 0.983 & 0.198 \\
 & $\alpha=0.25$ & 0.0841 $\pm$ 0.0004 & 0.0480 & 1624 $\pm$ 27 & 0.895 & 0.539 & 0.0439 $\pm$ 0.0004 & 0.0224 & 552 $\pm$ 11 & 0.978 & 0.231 \\
\addlinespace[2pt]
 & cap at $\tau_{99}$ & 0.0840 $\pm$ 0.0004 & 0.0479 & 1130 $\pm$ 24 & 0.930 & 0.453 & 0.0434 $\pm$ 0.0005 & 0.0221 & 391 $\pm$ 14 & 0.985 & 0.178 \\
 & cap at $\tau_{90}$ & 0.0832 $\pm$ 0.0003 & 0.0479 & 1531 $\pm$ 13 & 0.904 & 0.514 & 0.0429 $\pm$ 0.0002 & 0.0205 & 511 $\pm$ 9 & 0.979 & 0.219 \\
 & cap at $\tau_{50}$ & 0.0818 $\pm$ 0.0004 & 0.0468 & 1769 $\pm$ 27 & 0.883 & 0.562 & 0.0427 $\pm$ 0.0001 & 0.0208 & 656 $\pm$ 18 & 0.973 & 0.255 \\
\addlinespace[2pt]
 & ramp over 200 steps & 0.0841 $\pm$ 0.0002 & 0.0484 & 1529 $\pm$ 5 & 0.901 & 0.527 & 0.0450 $\pm$ 0.0004 & 0.0228 & 543 $\pm$ 5 & 0.978 & 0.230 \\
 & ramp over 400 steps & 0.0848 $\pm$ 0.0005 & 0.0486 & 1223 $\pm$ 23 & 0.923 & 0.475 & 0.0458 $\pm$ 0.0002 & 0.0232 & 345 $\pm$ 15 & 0.984 & 0.193 \\
\addlinespace[2pt]
 & $r_{\mathrm{std}}$, floor $\hat\sigma_{75\%}$ & 0.0814 $\pm$ 0.0000 & 0.0460 & 1860 $\pm$ 8 & 0.876 & 0.581 & 0.0468 $\pm$ 0.0006 & 0.0240 & 1167 $\pm$ 31 & 0.951 & 0.352 \\
 & $r_{\mathrm{std}}$, floor $\hat\sigma_{50\%}$ & 0.0816 $\pm$ 0.0004 & 0.0461 & 1891 $\pm$ 5 & 0.874 & 0.587 & 0.0468 $\pm$ 0.0002 & 0.0242 & 1438 $\pm$ 35 & 0.940 & 0.398 \\
 & $r_{\mathrm{std}}$, floor $\hat\sigma_{25\%}$ & 0.0822 $\pm$ 0.0005 & 0.0462 & 1888 $\pm$ 20 & 0.874 & 0.585 & 0.0464 $\pm$ 0.0004 & 0.0238 & 1575 $\pm$ 27 & 0.933 & 0.423 \\
 & $r_{\mathrm{std}}$, floor $\hat\sigma_{10\%}$ (main text) & 0.0818 $\pm$ 0.0001 & 0.0460 & 1915 $\pm$ 15 & 0.872 & 0.593 & 0.0464 $\pm$ 0.0002 & 0.0238 & 1652 $\pm$ 50 & 0.930 & 0.436 \\
 & $r_{\mathrm{std}}$, no floor & 0.0819 $\pm$ 0.0003 & 0.0461 & 1918 $\pm$ 31 & 0.872 & 0.593 & 0.0462 $\pm$ 0.0002 & 0.0237 & 1709 $\pm$ 29 & 0.928 & 0.444 \\
\midrule
\multirow{14}{*}{Atomic IDs} & $\alpha=2$ (mean doubled) & 0.0871 $\pm$ 0.0001 & 0.0471 & 479 $\pm$ 3 & 0.969 & 0.209 & 0.0486 $\pm$ 0.0006 & 0.0249 & 145 $\pm$ 2 & 0.991 & 0.077 \\
 & $\alpha=0.75$ & 0.0894 $\pm$ 0.0001 & 0.0485 & 597 $\pm$ 2 & 0.959 & 0.250 & 0.0495 $\pm$ 0.0002 & 0.0262 & 186 $\pm$ 5 & 0.989 & 0.086 \\
 & $\alpha=0.5$ & 0.0895 $\pm$ 0.0001 & 0.0487 & 663 $\pm$ 3 & 0.955 & 0.267 & 0.0501 $\pm$ 0.0004 & 0.0267 & 205 $\pm$ 4 & 0.988 & 0.091 \\
 & $\alpha=0.25$ & 0.0904 $\pm$ 0.0003 & 0.0492 & 765 $\pm$ 10 & 0.949 & 0.293 & 0.0507 $\pm$ 0.0002 & 0.0271 & 232 $\pm$ 2 & 0.987 & 0.097 \\
\addlinespace[2pt]
 & cap at $\tau_{99}$ & 0.0894 $\pm$ 0.0002 & 0.0484 & 622 $\pm$ 3 & 0.960 & 0.247 & 0.0503 $\pm$ 0.0001 & 0.0267 & 212 $\pm$ 2 & 0.988 & 0.088 \\
 & cap at $\tau_{90}$ & 0.0899 $\pm$ 0.0002 & 0.0490 & 722 $\pm$ 1 & 0.953 & 0.272 & 0.0505 $\pm$ 0.0004 & 0.0268 & 243 $\pm$ 2 & 0.986 & 0.096 \\
 & cap at $\tau_{50}$ & 0.0907 $\pm$ 0.0005 & 0.0493 & 802 $\pm$ 6 & 0.946 & 0.304 & 0.0505 $\pm$ 0.0001 & 0.0269 & 265 $\pm$ 1 & 0.985 & 0.106 \\
\addlinespace[2pt]
 & ramp over 200 steps & 0.0896 $\pm$ 0.0002 & 0.0488 & 662 $\pm$ 2 & 0.955 & 0.267 & 0.0501 $\pm$ 0.0003 & 0.0267 & 206 $\pm$ 1 & 0.988 & 0.092 \\
 & ramp over 400 steps & 0.0895 $\pm$ 0.0002 & 0.0486 & 602 $\pm$ 10 & 0.959 & 0.252 & 0.0496 $\pm$ 0.0004 & 0.0263 & 187 $\pm$ 8 & 0.989 & 0.088 \\
\addlinespace[2pt]
 & $r_{\mathrm{std}}$, floor $\hat\sigma_{75\%}$ & 0.0904 $\pm$ 0.0000 & 0.0494 & 864 $\pm$ 6 & 0.941 & 0.323 & 0.0510 $\pm$ 0.0003 & 0.0272 & 308 $\pm$ 4 & 0.983 & 0.116 \\
 & $r_{\mathrm{std}}$, floor $\hat\sigma_{50\%}$ & 0.0902 $\pm$ 0.0001 & 0.0493 & 867 $\pm$ 5 & 0.941 & 0.324 & 0.0516 $\pm$ 0.0003 & 0.0274 & 335 $\pm$ 7 & 0.981 & 0.124 \\
 & $r_{\mathrm{std}}$, floor $\hat\sigma_{25\%}$ & 0.0903 $\pm$ 0.0004 & 0.0493 & 865 $\pm$ 7 & 0.940 & 0.325 & 0.0513 $\pm$ 0.0001 & 0.0273 & 357 $\pm$ 2 & 0.980 & 0.128 \\
 & $r_{\mathrm{std}}$, floor $\hat\sigma_{10\%}$ (main text) & 0.0903 $\pm$ 0.0007 & 0.0493 & 874 $\pm$ 10 & 0.940 & 0.326 & 0.0513 $\pm$ 0.0002 & 0.0273 & 366 $\pm$ 3 & 0.980 & 0.130 \\
 & $r_{\mathrm{std}}$, no floor & 0.0904 $\pm$ 0.0005 & 0.0493 & 874 $\pm$ 13 & 0.940 & 0.326 & 0.0513 $\pm$ 0.0001 & 0.0273 & 369 $\pm$ 5 & 0.980 & 0.130 \\
\bottomrule
\end{tabular*}

\end{table}

\begin{multicols}{2}
\paragraph{\textbf{Paired tests.}} \looseness=-1 Every RL arm is scored on the same test users as
the SFT checkpoint it resumes from, so HR@10 is compared per user with an
exact McNemar test and NDCG@10 with a paired bootstrap ($2000$
resamples), seed $s$ against seed $s$. Against the start, the CF arm
moves HR@10 by $-0.0006$ to $-0.0013$ on Beauty ($p \geq 0.33$) and
$-0.0052$ to $-0.0061$ on Sports ($p \leq 10^{-4}$) with Semantic IDs and
by at most $0.0017$ with Atomic IDs ($p \geq 0.19$), and NDCG@10 by
$-0.0017$ to $-0.0019$ on Beauty with Semantic IDs ($p \leq 0.015$). With
Semantic IDs on Beauty, per-item centering is below the CF arm in HR@10
by $0.0028$ to $0.0031$ ($p = 0.020$ to $0.041$) and below the start by
$0.0035$ to $0.0044$ ($p \leq 0.008$); per-item standardization is below
the CF arm by $0.0025$ ($p \geq 0.05$ on every seed) and below the start
by $0.0033$ ($p < 0.05$ on every seed). Elsewhere per-item centering is
above the CF arm in HR@10 on every seed and gains $0.0009$ to $0.0012$ in
NDCG@10 with Atomic IDs ($p \leq 0.04$); per-item standardization gains
$0.0046$ in HR@10 over the CF arm and $0.0030$ over per-item centering on
Sports with Semantic IDs ($p < 0.05$ on every seed). By tier
(Table~\ref{tab:tiers}), every CF seed is below the start in mid-tier
HR@10 ($p \leq 0.011$), and every seed of both corrections is above the CF
arm ($p \leq 0.021$ for centering, $p < 10^{-5}$ for standardization), in
all four settings; in the head tier both are below the CF arm on Beauty
only (standardization 0.1416 against 0.1556 with Semantic IDs,
$p < 0.01$).

\end{multicols}

\end{document}